\documentclass[useAMS,usenatbib]{mn2e}

\usepackage{graphicx, pstricks}

  \newcommand{\eg}{e.g.}

\usepackage[breaklinks]{hyperref}
\usepackage{amssymb}
\hypersetup{
    unicode=false,          
    pdftoolbar=true,        
    pdfmenubar=true,        
    pdffitwindow=false,     
    pdfauthor={P. Tanga et al.},
    pdfsubject={Planetary Science},
    pdfkeywords={},         
    pdfnewwindow=true,      
    colorlinks=true,        
    linkcolor=gray,         
    citecolor=blue,         
    filecolor=gray,         
    urlcolor=gray           
}

\usepackage{ulem}

\newcommand{\addb}[1]{\textbf{#1}}

\title[The non-convex shape of (234) Barbara, the first Barbarian]
  {The non-convex shape of (234) Barbara, the first Barbarian
    \thanks{Based in part on observations collected at the European Organization for
  Astronomical Research in the Southern Hemisphere, Chile - program
  ID: 
  \href{http://archive.eso.org/wdb/wdb/eso/eso_archive_main/query?prog_id=076.C-0798}{076.C-0798}.}}
  
\author[P. Tanga et al.]
  {\Large P.~Tanga,$^{1}$ 
  B.~Carry,$^{2}$ 
  F.~Colas,$^{2}$ 
  M.~Delbo,$^{1}$
  A.~Matter,$^{1,3}$
  J.~Hanu{\v s},$^{1,4}$
  V. Al\'i Lagoa,$^{1}$
  \large A.H.~Andrei,$^{5,9,29}$
  \newauthor
  \large M.~Assafin,$^{5}$
  M.~Audejean,$^{6,7}$
  R.~Behrend,$^{6,8}$
  J.I.B.~Camargo,$^{9}$
  A.~Carbognani,$^{10}$
  M.~Cedr{\'e}s Reyes,$^{11}$
  M.~Conjat,$^{12}$
  \newauthor
  \large N.~Cornero,$^{6,13}$
  D.~Coward,$^{14}$
  R.~Crippa,$^{6,15}$
  E.~de Ferra Fantin,$^{11,16}$
  M.~Devog{\'e}le,$^{17}$
  G. Dubos,$^{13}$
  E.~Frappa,$^{18}$
  M. Gillon,$^{17}$
  \newauthor
  \large H.~Hamanowa,$^{6,19}$
  E.~Jehin,$^{17}$
  A.~Klotz,$^{20}$
  A.~Kryszczy{\'n}ska ,$^{21}$
  J.~Lecacheux,$^{22}$
  A.~Leroy,$^{6,23}$
  J. Manfroid,$^{17}$
  F.~Manzini,$^{6,24}$
  \newauthor
  \large L.~Maquet,$^{2}$
  E.~Morelle,$^{6,25}$
  S.~Mottola,$^{26}$
  M.~Poli{\'n}ska,$^{21}$
  R.~Roy,$^{6,27}$
  M.~Todd,$^{14}$
  F.~Vachier,$^{2}$
  C.~Vera Hern{\'a}ndez,$^{11}$
  \newauthor
  \large P.~Wiggins$^{28}$
  \\
$^{1}${Laboratoire Lagrange, UMR7293 CNRS, UNS, Observatoire de la C\^ote d'Azur, Nice, France}\\
$^{2}${IMCCE, Observatoire de Paris, UMR8028 CNRS, France}
$^{3}$Max Planck Institut fÃŒr Radioastronomie, Bonn, Germany
$^{4}$Astronomical Institute,\\ Faculty of Mathematics and
 Physics, Charles University in Prague, Czech Republic
$^{5}$Observat{\'o}rio do Valongo/UFRJ, Brazil\\
$^6$CdR \& CdL Group: Lightcurves of Minor Planets and Variable Stars, Switzerland
$^{7}$Observatoire de Chinon, Chinon, France\\
$^{8}$Geneva Observatory, Switzerland
$^{9}$Observat{\'o}rio Nacional/MCTI, Brazil
$^{10}$Osservatorio Astronomico della regione autonoma Valle d'Aosta,\\ Italy
$^{11}$Agrupaci{\'o}n Astron{\'o}mica de Fuerteventura, Spain
$^{12}$Observatoire de Cabris, France
$^{13}$Association des Utilisateurs de\\ D{\'e}tecteurs {\'E}lectroniques (AUDE), France
$^{14}$Department of Imaging and Applied Physics, Curtin University of Technology, Bentley, Australia\\
$^{15}$Osservatorio astronomico di Tradate, Italy
$^{16}$Academia de ciencias e ingienerias de Lanzarote, Arrecife, Spain
$^{17}$Institut d'Astrophysique, \\G{\'e}ophysique et Oc{\'e}anographie, Universit{\'e} de Li{\`e}ge, Belgium
$^{18}$Euraster, St. Etienne, France
$^{19}$Hamanowa Astronomical Observatory, Fukushima, \\ Japan
$^{20}$CNRS, IRAP, Toulouse, France
$^{21}$Astronomical Observatory Inst., Faculty of Physics, Adam Mickiewicz University, Pozna{\'n}, Poland\\
$^{22}$LESIA-Observatoire de Paris, CNRS, UPMC Univ. Paris 06, Univ.  Paris-Diderot, Meudon, France
$^{23}$Association T60, Toulouse, France\\ 
$^{24}$Stazione Astronomica di Sozzago, Italy
$^{25}$Lauwin Planque, France
$^{26}$Institute of Planetary Research, German Aerospace Center, Berlin, \\ Germany
$^{27}$Blauvac Observatory, St.-Est{\`e}ve, France
$^{28}$Wiggins Observatory, Tooele, UT, USA
$^{29}$SYRTE, Observatoire de Paris, France
  }
\pagerange{\pageref{firstpage}--\pageref{lastpage}} \pubyear{2015}
\def\LaTeX{L\kern-.36em\raise.3ex\hbox{a}\kern-.15em
    T\kern-.1667em\lower.7ex\hbox{E}\kern-.125emX}

\begin{document}
\label{firstpage}
\maketitle
\begin{abstract}
  Asteroid (234) Barbara is the prototype of a category of asteroids that has been shown to
  be extremely rich in refractory inclusions, the oldest material ever
  found in the Solar System. It exhibits several peculiar features,
  most notably its polarimetric behavior. In recent years other
  objects sharing the same property (collectively known as ''Barbarians'') have been discovered.
  Interferometric observations
    in the mid-infrared with the ESO VLTI suggested that
  (234) Barbara might have a bi-lobated shape or even a large
  companion satellite.
We use a large set of 57 optical lightcurves acquired between
    1979 and 2014, together with the timings of two stellar
    occultations in 2009, to determine the rotation period, spin-vector
    coordinates, and 3-D shape of (234) Barbara, using two different shape
    reconstruction algorithms.
By using the lightcurves combined to
  the results obtained from stellar occultations, we are able to show that the shape of (234) Barbara exhibits large concave areas. 
  Possible links of the shape to the polarimetric properties and the object evolution are discussed. We also show that VLTI data can
  be modeled without the presence of a satellite.
\end{abstract}

\section{Introduction}
\label{Intro}

  The physical characterization of asteroids is of primary importance
  for understanding their origin and evolution. Simple information
  such as rotation period and direction of the spin axis have been
  related to evolutionary processes such as
  accretion in the protoplanetary disk \citep{Johansenetal2010}, 
  impacts \citep{Takeda2009_I, Takeda2009_II, Marzari11,
    holsapple_momentum_2012},  thermal forces
  \citep{bottke_yarkovsky_2006}, internal cohesion and degree of
  fragmentation \citep{holsapple07}, to cite a few notable examples. 
  
  Asteroid (234) Barbara exhibits peculiar features, such as an anomalous
  polarimetric behavior \citep{cellino_strange_2006}, a possible very irregular shape, a suspected
  large companion \citep{delbo_VLTI_2009}, a long rotation period (Schober 1981), as detailed in
  Sect.~\ref{S:Barbara}. Polarimetry and near-infrared spectroscopy suggest
  that (234) Barbara and other similar asteroids could be composed by
  high fraction of the most ancient solids formed in the Solar
  System, the Ca-Al-rich Inclusions \citep[CAI, see][]{sunshine08}.  

  These properties motivated a focused study of this target, given the
  substantial lack of other physical data. 
  Here we address in particular the determination of the shape and the   infirmation of the binary hypothesis. We proceeded
  by setting up a long and intense campaign of photometric
  observations, complemented with two stellar occultations.
 
Traditional lightcurve inversion
  \citep[\eg,][]{2001-Icarus-153-Kaasalainen-a,
    2001-Icarus-153-Kaasalainen-b}, retrieving complex shapes
  described by several parameters, converge to a unique solution only
  under the hypothesis of convexity  of the shape. In the specific case of (234) Barbara, an imposed convexity could hide the evidence of a bi-lobated structure,
  responsible of the suspected presence of a satellite
\citep{delbo_VLTI_2009}.
     
  Given the limitations of photometry when taken alone, we also
 apply the inversion algorithm KOALA
  \citep{2010-Icarus-205-Carry-a, 2011-IPI-5-Kaasalainen}, which can
  use data coming from different sources for deriving a consistent,
  unique model of an asteroid. Its main applications have concerned
  the joint inversion of photometry, disk-resolved imaging and
  stellar occultations. We illustrate in the following the results
  that we obtained on the asteroid (234) Barbara by this approach. 

  Our efforts were focused on an observation campaign in the period
  2008-2011 (plus some additional data in 2014), involving both time-series photometry and stellar occultations, whose results are presented in
  Sect.~\ref{S:photom} and~\ref{S:occult}. We were able to invert both
  photometric and occultation data for deriving a 3-D shape model, as
  described in Sect.~\ref{S:LC_inversion}. 
  We use this model to validate the interferometric observations at VLTI
 presented by \citet{delbo_VLTI_2009}, and eventually discuss the implications of our results.

\section{Peculiarities of (234) Barbara}
\label{S:Barbara}

  \indent Barbara is an asteroid belonging to the inner Main Belt,
  classified for a long time as a S type (Tedesco 1989). Its slow
  rotation (about 26.5 hours) was discovered by Schober (1981). 
  A detailed, dedicated, physical characterization for this object was
  not attempted in the past, with the exception of spectroscopy. Owing to a slight excess in reflectance in the red
  part of the spectrum with respect to the core of the S class,
  followed by a flat plateau in the near-infrared, Barbara was
  classified Ld in the \citet{bus_binzel_2_2002} taxonomy. 
  From thermal radiometry with IRAS, a diameter of 44\,$\pm$\,1\,km
  was derived 
  \citep{2002-AJ-123-Tedesco-a} corresponding to a geometric albedo
  $p_v$\,=\,0.22\,$\pm$\,0.01 (assuming an absolute magnitude of
  $H=9.02$). Other available size determinations involve AKARI (47.8$\pm$0.68 km) and the Wide Infrared Survey Explorer (WISE). WISE yielded two measurements, but one of them is clearly discrepant. We discuss this issue in detail and provide a new, coherent diameter determination, in Sect.\ref{S:LC_inversion}.

  \citet{cellino_strange_2006} pointed out that this asteroid has an
  unusual polarimetric behavior. 
  The degree of linear polarization of sunlight scattered by asteroid
  surfaces exhibits a variation as a function of the illumination
  conditions, described by means of the phase angle, namely the angle
  between the directions to the Sun and to the observer, as seen from
  the asteroid. In particular, the morphology of the
  phase-polarization curve has some general properties which, apart
  from some differences related mainly to the geometric albedo of the
  surface, tend to be shared by all known asteroids. The case of (234)
  Barbara is different, as it exhibits a ``negative polarization
  branch'' wider than usual, with an ``inversion angle'' around
  30$^\circ$, a much larger value with respect to the $\sim$20$^\circ$
  displayed by other objects (for details see
  \citealt{cellino_strange_2006}). We recall here that the
  negative polarization corresponds to a polarization plane parallel to
  the scattering plane.  

  A very similar phase-polarization curve was found later on for other
  L, Ld and K-type asteroids  \citep{gil-hutton_new_2008,
    masiero_polarization_2009}, collectively known as ``Barbarians''
  from the first discovered. Data concerning the asteroid (21) Lutetia
  seems to indicate also a peculiar polarization, but with a lower
  inversion angle \citep{2010-AA-515-Belskaya} intermediate between  
  \textsl{regular} asteroids and \textsl{Barbarians}.

  Several hypothesis have been formulated in the past for explaining the
  high fraction of negative polarization. In particular, the role of
  coherent backscattering \citep{Muinonen89, Shkuratov94} was invoked,
  normally associated to high albedos producing narrow and strong 
  opposition peaks in the phase-brightness curve. Another possibility
  is single particle scattering \citep{Munoz00, Shkuratov02} on refractory inclusions. In fact, among
  meteorites, some carbonaceous chondrites (type CV3 and CO3) produce
  the highest negative polarization, which could be related to the
  abundance of fine-grained Ca-Al refractory inclusions (CAI)
  \citep{Zellner77}. \citet{Burbine92} suggested that (980) Anacostia,
  another object having polarimetric properties similar to (234)
  Barbara, could be a spinel-rich body with a mineralogy similar 
  to CO3/CV3 meteorites. The reason of the negative polarization should then
  be related to the fine-grained structures of white spinel inclusions
  surrounded by a dark matrix \citep{Burbine92}. \citet{sunshine08}
  reached similar conclusions for Barbara itself by comparing its
  visible and IR spectra to laboratory spectra of CAI
  materials. Surprisingly, a satisfactory match can be reached only
  when the fraction of spinel-bearing CAIs is very high (up to 30\%
  for explaining the spectral features observed). If this finding were
  true, the Barbarians should have formed in a nebula rich of
  refractory materials, and would be among the most ancient asteroids
  formed. No sample with high percentages of CAI is present in the
  current meteorite collections. \\
  \indent If mineralogy is the culprit for the polarization anomaly, it is
  then not surprising that all Barbarians belong to a similar spectral
  class. At first sight, it is unclear why not all other L, Ld, and
  K asteroids (and more in general the whole S-complex) do not share
  similar polarization properties. For example the L-type (12)
  Victoria has a usual polarization with inversion angle
  $\sim$20\degr. However, if the near-IR spectrum is considered, all Barbarians belong to 
  the same L class as defined by (De Meo et al. 2009). \\ 
  \indent \citet{cellino_strange_2006} suggested that
  anomalous polarization could be due to large-scale concavities responsible of introducing a distribution of scattering and incidence angles different from those of a regular, convex surface. This conjecture has never been proven to play a role, essentially due to the lack of measurements at high phase
  angles of objects known to have large concave features. Also, theoretical models do not seem to explore explicitly this possibility.\\
  Asteroid (234) Barbara was one of the targets chosen for the first
  interferometric observations of asteroids by the Very Large
  Telescope Interferometer (VLTI), using  the MID-infrared
  Interferometer \citep[MIDI,][]{2003-ApSS-286-Leinert}. 
  This instrument can reach an angular resolution in the range 20-200
  mas, depending on the choice of the baseline. It can thus be used to
  measure the apparent size of asteroids, 
  by modeling the visibility of the interferometric fringes,
  independently from other common methods, such as thermal infrared
  radiometry. It can also be used to detect and study the orbit of
  multiple asteroid systems. VLTI-MIDI observations, obtained in
  November 2005, yielded an average diameter of
  44.6\,$\pm$\,0.3\,km. These observations and their processing are
  extensively described in \citet{delbo_VLTI_2009}. Their main result, obtained by modeling the VLTI visibility function, is the detection of a signature of duplicity of the source. In fact, they show that a fit with a single object cannot reproduce the data. A satisfactory fit requires a second component. The resulting system, composed by two uniform disks
  $\sim$37 and $\sim$21\,km in diameter, could be interpreted
  either as a single object of irregular shape, or as a binary close to the alignment with the line of sight. This model is reproduced for
  ease of comparison in Fig.~\ref{F:VLTI}. 
  
  The unambiguous hints of a binary or bi--lobated body has been an
  additional motivation for our focused study of (234) Barbara. We stress here that we don't use VLTI data in the shape determination process described below (a rather difficult task to the very different nature of such data). Conversely, we will test our derived shape against the VLTI observations.
  
  Very recently, by observing the polarization properties of the
  members of the high-inclination dynamical family of the L-type (729) Watsonia, \citet{Cellinoetal2014} discovered that it is composed by Barbarians.
  This fact seems to indicate that the
  Barbarian character is intrinsic to bulk properties of the body, not
  only to some surface effects. In fact, if the anomalous polarization
  was limited to a surface process, the parent body shattering and the
  subsequent fragment mixing, would strongly dilute the polarimetric
  signature of the original surface on the family members. 
  
\section{Observations\label{S:obs}}

  \subsection{Photometric campaign\label{S:photom}}

    We obtained R-band photometry of Barbara starting in June 2008. The
    most recent observations that we use in our data reduction have been
    acquired in February 2014. Given the long rotation period, a single
    site is highly inefficient in covering a full rotation. We
    thus put at contribution many observers and telescopes, at widely
    different longitudes. Thanks to this considerable, shared effort we were able to
    provide an adequate coverage of the brightness variations.  

\begin{figure} 
      \centerline{\includegraphics[width=0.49\textwidth]{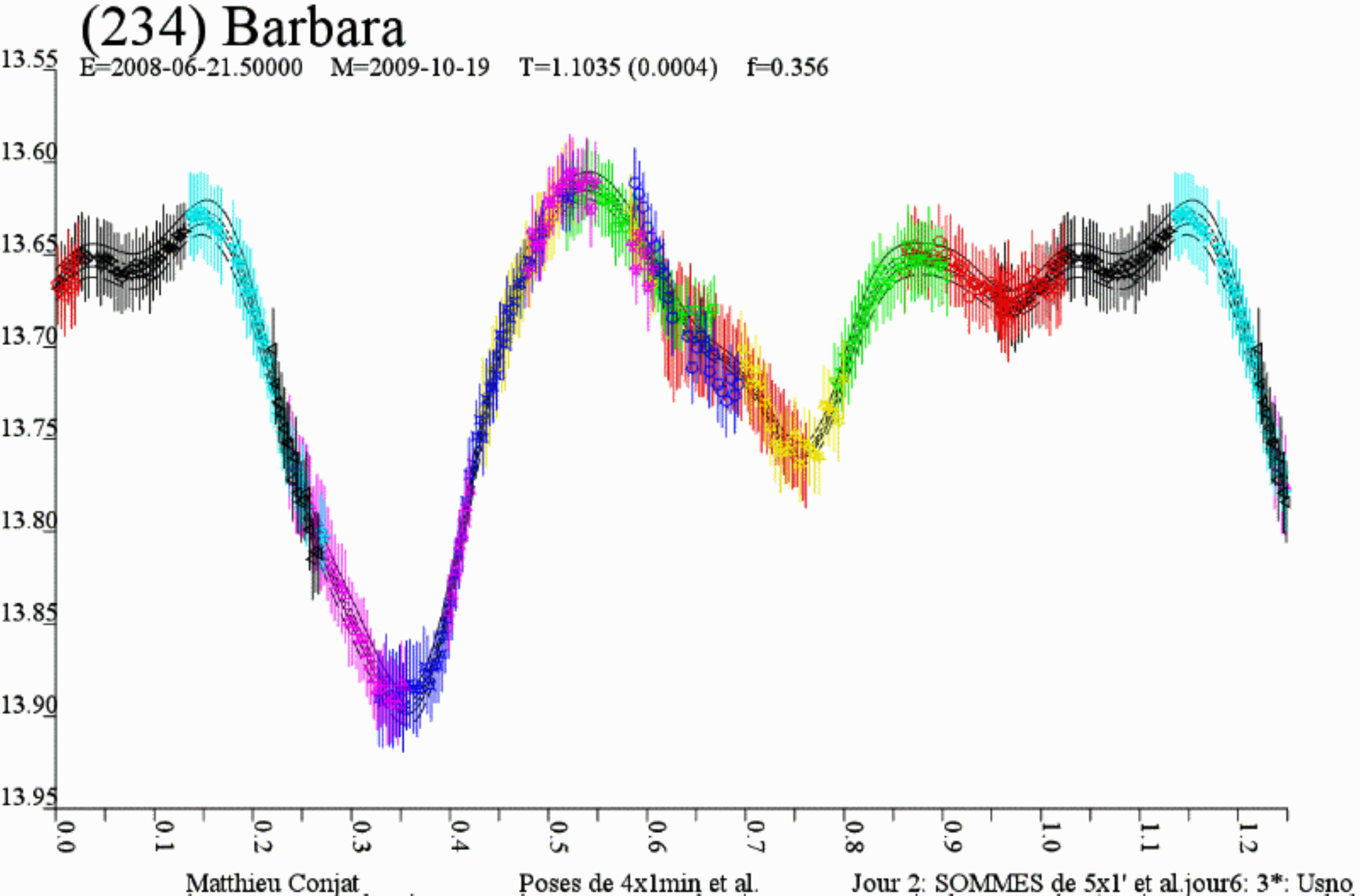}}
      \caption{
        Lightcurve obtained in June 2008 over 13 days by M. Conjat (from the main OCA site) and P. Tanga (Tourrette-Levens, private facility), folded on
          the rotation period of 26.474\,h. Each color represents a
          different observing session.
       As the observing sites were in the same geographic
        area, several nights were needed to cover the entire
        rotation. The curve qualitatively represent the amplitude and
        shape of the brightness variation, hinting to an irregular
        body. 
      }
      \label{F:Conjat_Tanga}
    \end{figure}

    In Table~\ref{T:photometry} we list the different photometric data
    sets that we used for this study. A sample lightcurve is
    presented in Fig.~\ref{F:Conjat_Tanga}. This result is the composite of 13 observing sessions spanning $\sim 6$ weeks, from two sites very close in longitude. Over this time, the object 
    changes its geometry, relative to the Sun and to the observer, thus introducing 
    potentially complex variations that are not linked to the object
    rotation alone. In particular, phase angle variations can rapidly change shadowing effects.
    For this reason the folded lightcurve has just
    an indicative value. From a qualitative point of view, it is
    however interesting to note that the amplitude is rather large and
    the variation complex. 

\begin{table*}\scriptsize
\centering
\caption[]{Photometric observing runs used to derive the shape of (234) Barbara. The full table is available among the online resources. Legend: $\aleph$: 0.32 m Observatoire du Chinon (France): M. Audejean.
$\flat$: 0.40 m, Hamanowa Astronomical Observatory (Japan): H. \& H. Hamanowa.
$\Box$: 0.4 m Observatoire de la C\^ote d'Azur (France): M. Conjat; and 0.2 m Specola Tourrette Levens: P. Tanga.
$\uplus$: From Schober (1981).
\label{T:photometry}			           		             
} 
\begin{tabular}{rllcrrl}
\hline
\multicolumn{4}{l}{Session start (yyyy-mm-dd hh:mm)} & Duration (min.) & N obs. & Notes \\ 
\hline		
1979& 09& 13&  05:11 & 262   &      138 &      $\uplus$  \\
1979& 09& 14&  01:54 & 455   &      212 &      $\uplus$  \\
1979& 09& 15&  02:06 & 437   &       64 &      $\uplus$  \\
2008& 06& 18&  20:56 & 226   &       51 &      $\Box$   \\
2008& 06& 19&  20:46 & 236   &       41 &      $\Box$  \\
2008& 06& 20&  20:58 & 279   &       34 &      $\Box$ \\
2008& 06& 21&  20:41 & 313   &       52 &      $\Box$ \\
2008& 06& 22&  20:45 & 291   &       41 &      $\Box$ \\
2008& 06& 23&  20:35 & 216   &       34 &      $\Box$  \\
2008& 06& 25&  20:41 & 283   &       50 &      $\Box$ \\
2008& 06& 26&  20:45 & 256   &       46 &      $\Box$ \\
2008& 06& 27&  21:21 & 182   &       41 &      $\Box$ \\
2008& 06& 28&  21:18 & 157   &       36 &      $\Box$\\
2009& 11& 18&  23:21 & 22 & 51 & $\Im$ \\
2009& 11& 24&  23:16 & 121   &	   25 &	     $\aleph$\\
2009& 11& 25&  22:55 & 160   &       13 &	     $\aleph$\\
2009& 11& 25&  12:41 & 206   &	   17 &	     $\flat$\\
... & & & & & & \\
\multicolumn{7}{l}{Complete table available in the online resources.}
\end{tabular}
\end{table*}


    We can judge the long-term photometric coverage by considering the
    ranges of ecliptic heliocentric longitudes corresponding to the
    three apparitions over which the asteroid was observed. We thus
    obtain $\sim$357\degr~in September 1997,
    260\degr~in June 2008,
    160\degr-180\degr~in the period November 2010 - April 2011,
      and 120\degr in February 2014.
    The distribution over different values in ecliptic
    longitude favors the coverage of the largest possible range of
    aspects angle allowed by the pole obliquity.  
    
  \subsection{Stellar occultations\label{S:occult}}

    \indent Two stellar occultations by (234) Barbara were successfully observed by our team
    and collaborators, on October 5, 2009 and only a few weeks later, on November 21, 2009. 
    Both target stars had a magnitude V$<8$, which greatly facilitated the use of portable 
    instruments and the deployment of several stations.
    
    \indent The first event took place on the Atlantic Ocean, with a predicted path passing on Canary Islands
    and central Africa. An expedition from France installed portable observing stations on the 
    islands of Fuerteventura and Lanzarote (southern tip). Local amateur astronomers contributed
    with equipment at further sites, and logistic support. Good weather granted nearly optimal conditions and 
    several chords of the occultation were recorded (see
    Table~\ref{T:occ1}). The positive result showed that the
    prediction was very accurate, the real shadow being shifted
    southward relatively to predictions by $\sim$10\,km only.  
    
    The high reliability of the asteroid positional ephemeris was a
    precious information for planning the deployment of the observers
    around the path of the following event. Several astronomers in the
    United States gathered in Florida and few others in central Europe
    at the oriental extreme of the path. The noticeable effort for the
    deployment of several automated stations by single observers was
    successful, thus securing a dense set of occultation chords
    (Table~\ref{T:occ2}). For both events the entire data set is
    available, for example, at
    PDS\footnote{\addb{\href{http://sbn.psi.edu/pds/resource/occ.html}{http://sbn.psi.edu/pds/resource/occ.html}}} \citep{PDSSBN-OCC}.  
    
\begin{figure}
\centering
  \includegraphics[width=0.49\textwidth]{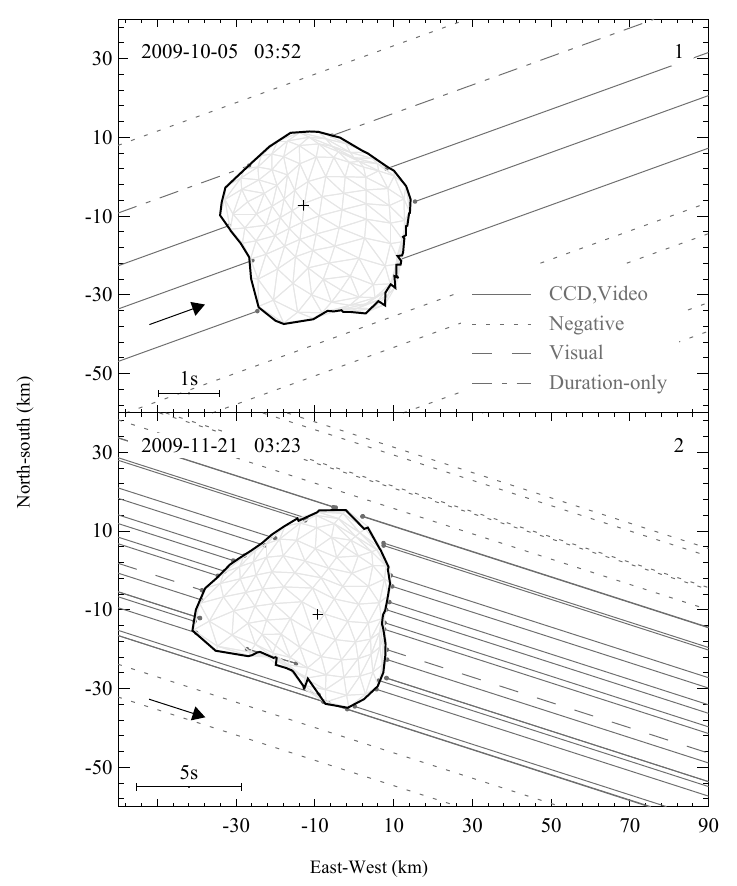}%
   \caption[Stellar occultations]{%
    Plot on the plane of the sky of the occultation chords reported in
    Tables~\ref{T:occ1} and~\ref{T:occ2}. The different line formats distinguish among positive observations (CCD, Video), negative ones, visual timings, and duration measurements. 
    The profile of (234) Barbara at the epoch of each
    occultation, as derived by the shape model presented in this
    article, is represented by the black contours. 
    \label{F:occ}
  }
\end{figure}

    Data reduction of the occultation on October 5, 2009 event results in the strong hint of
    an elongated, oval shape (Fig.~\ref{F:occ}), with possible irregular features.
    The observed chords on November 21, 2009 consistently draw an overall triangular shape, 
    with a large and pronounced concavity at the South limb, and
    hints of other minor concavities (Fig.~\ref{F:occ}).
    Both occultations provide an average size consistent with the results obtained at VLTI 
    (Sect.~\ref{S:LC_inversion}).

    Other occultation events by Barbara were observed later on. We
    neglect here the results obtained on December 14, 2009 in the USA,
    given that only 3 chords are available \citep{PDSSBN-OCC}. Also,
    positive observations of the occultation on January 17, 2010,
    obtained in 
    Japan, have not been used as the faintness of the target star (V=12.0) prevented accurate timings. 
    
    None of the observed occultations presents secondary events linked
    to the presence of possible satellites in proximity of the primary
    body.  

\begin{table*}\scriptsize
\centering
  \caption[Stellar occultation of October 5, 2009]{%
    Observers and chords for the occultation of HIP32822 (V=7.77),
    on 2009 October 5. The UT columns contain the epochs associated to
    each chord. In the ``Event'' column,  
    the flags can be: M = missed, no occultation observed; D =
    disappearance; and R = reappearance.
    Uncertainties on timing in seconds at the 3-$\sigma$ level are
    provided.
    Some observers were capable of deploying multiple stations, so their
    name appear more than once.
    \label{T:occ1}		                	           
  }
\begin{tabular}{rl ccc ccc}
\hline                                                                                                       
\# & Observer and site & UT & Event & 3$\sigma$ & UT & Event & 3$\sigma$ \\
\hline  
 1 & Vachier, Morro del Jable, Spain      &             & M &      &             & M &      \\
 2 & Vachier, Costa Calma, Spain            &             & M &      &             & M &      \\
 3 & Lecacheux, Gran Tarajal, Spain      &             & M &      &             & M &      \\
 4 & Maquet, Tuineje, Spain               &             & M &      &             & M &      \\
 5 & Lecacheux, Antigua, Spain            & 04 10 22.02 & D & 0.03 & 04 10 24.50 & R & 0.03 \\
 6 & Maquet/de Ferra/Cedr\'es, Tefia, Spain & 04 10 22.25 & D & 0.02 & 04 10 25.11 & R & 0.02 \\
 7 & Colas/Vera, La Oliva, Spain          & 04 10 22.44 & D & 0.02 & 04 10 25.21 & R & 0.02 \\
 8 & Colas, Corralejo, Spain              & 04 10 23.50 & D & 0.15 & 04 10 24.96 & R & 0.15 \\
 9 & Tanga, Playa Blanca, Spain           &             & M &      &             & M &      \\
\hline					            	          		  	    
\end{tabular}			                	           
\end{table*}

\begin{table*}\scriptsize
\centering
\caption[Stellar occultation of November 2009]{%
  Similar as Table~\ref{T:occ1}, for the occultation of the star HIP34106
  (V=7.5), on 2009 November 21.
\label{T:occ2}			           		             
} 
\begin{tabular}{rl ccc ccc}
\hline
\# & Observer and site & UT & Event & 3$\sigma$ & UT & Event & 3$\sigma$ \\
\hline
 1 & Harris, Deltona, FL, USA              &             & M &      &             & M &      \\
 2 & Dunham, Okahumpka, FL, USA            &             & M &      &             & M &      \\
 3 & Dunham, Center Hill, FL, USA          &             & M &      &             & M &      \\
 4 & Venable, Webster, FL, USA   	        &             & M &      &             & M &      \\
 5 & Venable, Tarrytown, FL, USA           &             & M &      &             & M &      \\
 6 & Venable, Tarrytown, FL, USA           & 03 38 34.27 & D & 0.03 & 03 38 35.62 & R & 0.03 \\
 7 & Dunham, Groveland, FL, USA            & 03 38 32.77 & D & 0.10 & 03 38 34.21 & R & 0.10 \\
 8 & Maley, Clermont, FL, USA              & 03 38 32.18 & D & 0.05 & 03 38 36.01 & R & 0.05 \\
 9 & Fernandez/N Lust, Orlando, FL, USA    & 03 38 26.70 & D & 0.50 & 03 38 30.60 & R & 0.50 \\
10 & Dunham, Green Pond, FL, USA           & 03 38 32.63 & D & 0.02 & 03 38 38.37 & R & 0.02 \\
11 & Maley, Polk City, FL, USA             & 03 38 32.82 & D & 0.02 & 03 38 39.16 & R & 0.02 \\
12 & Turcani, Christmas, FL, USA           & 03 38 27.30 & D & 0.10 & 03 38 34.50 & R & 0.10 \\
13 & Bredner, Germany                      & 03 18 12.40 & D &  --  & 03 18 18.70 & R &  --  \\
14 & Maley, Polk City, FL, USA             & 03 38 33.90 & D & 0.10 & 03 38 41.50 & R & 0.02 \\
15 & Maley, Polk City, FL, USA             & 03 38 34.52 & D & 0.02 & 03 38 42.79 & R & 0.02 \\
16 & Povenmire, Deer Park, FL, USA         & 03 38 28.30 & D & 0.30 & 03 38 37.50 & R & 0.30 \\
17 & Denzau, Panker, Germany               & 03 18 02.68 & D &  --  & 03 18 10.64 & R &  --  \\
18 & Maley, Polk City, FL, USA             & 03 38 34.61 & D & 0.02 & 03 38 43.85 & R & 0.02 \\
19 & Iverson, Harmony, FL, USA             & 03 38 30.95 & D & 0.02 & 03 38 40.23 & R & 0.02 \\
20 & Coles, Harmony, FL, USA               & 03 38 30.73 & D & 0.03 & 03 38 40.01 & R & 0.03 \\
21 & Degenhardt, Deer Park, FL, USA        & 03 38 29.61 & D & 0.02 & 03 38 38.77 & R & 0.02 \\
22 & Degenhardt, Deer Park, FL, USA        & 03 38 36.26 & D & 0.02 & 03 38 38.79 & R & 0.02 \\
23 & Degenhardt, Deer Park, FL, USA        & 03 38 29.96 & D & 0.02 & 03 38 33.96 & R & 0.02 \\
24 & Degenhardt, Deer Park, FL, USA        & 03 38 31.70 & D & 0.10 & 03 38 33.50 & R & 0.02 \\
25 & Degenhardt, Deer Park, FL, USA        & 03 38 32.23 & D & 0.02 & 03 38 33.16 & R & 0.02 \\
26 & Degenhardt, Deer Park, FL, USA        & 03 38 33.27 & D & 0.02 & 03 38 33.40 & R & 0.02 \\
27 & Degenhardt, Deer Park, FL, USA        &             & M &      &             & M &      \\
28 & Bulder, Buinerveen, The Netherlands   &             & M &      &             & M &      \\
\hline					           		             
\end{tabular}				           		             
\end{table*}

\section{Shape and spin determination of (234) Barbara}
\label{S:LC_inversion}

  As the occultations indicate the clear presence of concavities and
  the lightcurve is rather irregular, a simultaneous inversion by
  KOALA of photometry and occultation data appeared necessary
  \citep[see][for a description of the
    algorithm]{2011-IPI-5-Kaasalainen, 2010-Icarus-205-Carry-a,
    2012-PSS--Carry}. For a further consistency check we also ran the
  usual lightcurve-only inversion
  \citep[\eg,][]{2001-Icarus-153-Kaasalainen-a,
    2001-Icarus-153-Kaasalainen-b} to retrieve a convex model, the
  close envelope of the concave shape. 

\begin{figure*}
    \centerline{\includegraphics[width=0.9\textwidth]{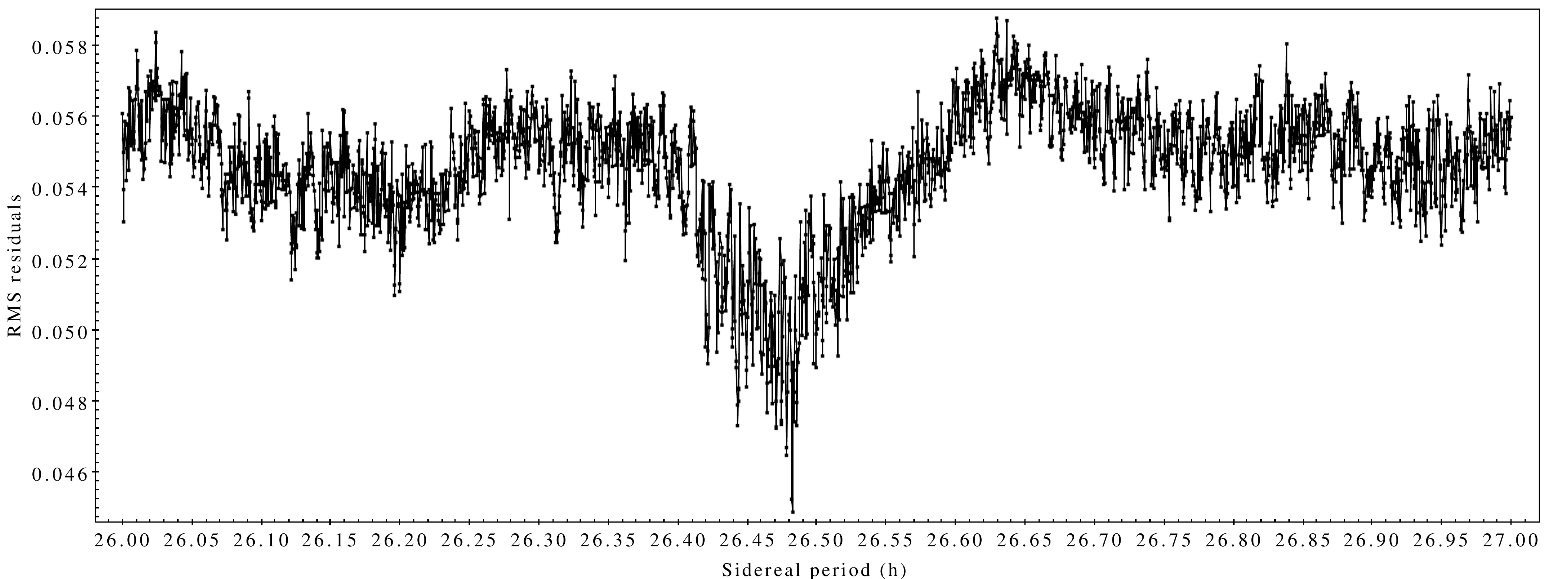}}
    \caption{The RMS deviation of the photometry relative to the
      convex model, against sidereal rotation period.} 
    \label{F:period}
  \end{figure*}

\begin{figure}
    \centerline{\includegraphics[width=0.49\textwidth]{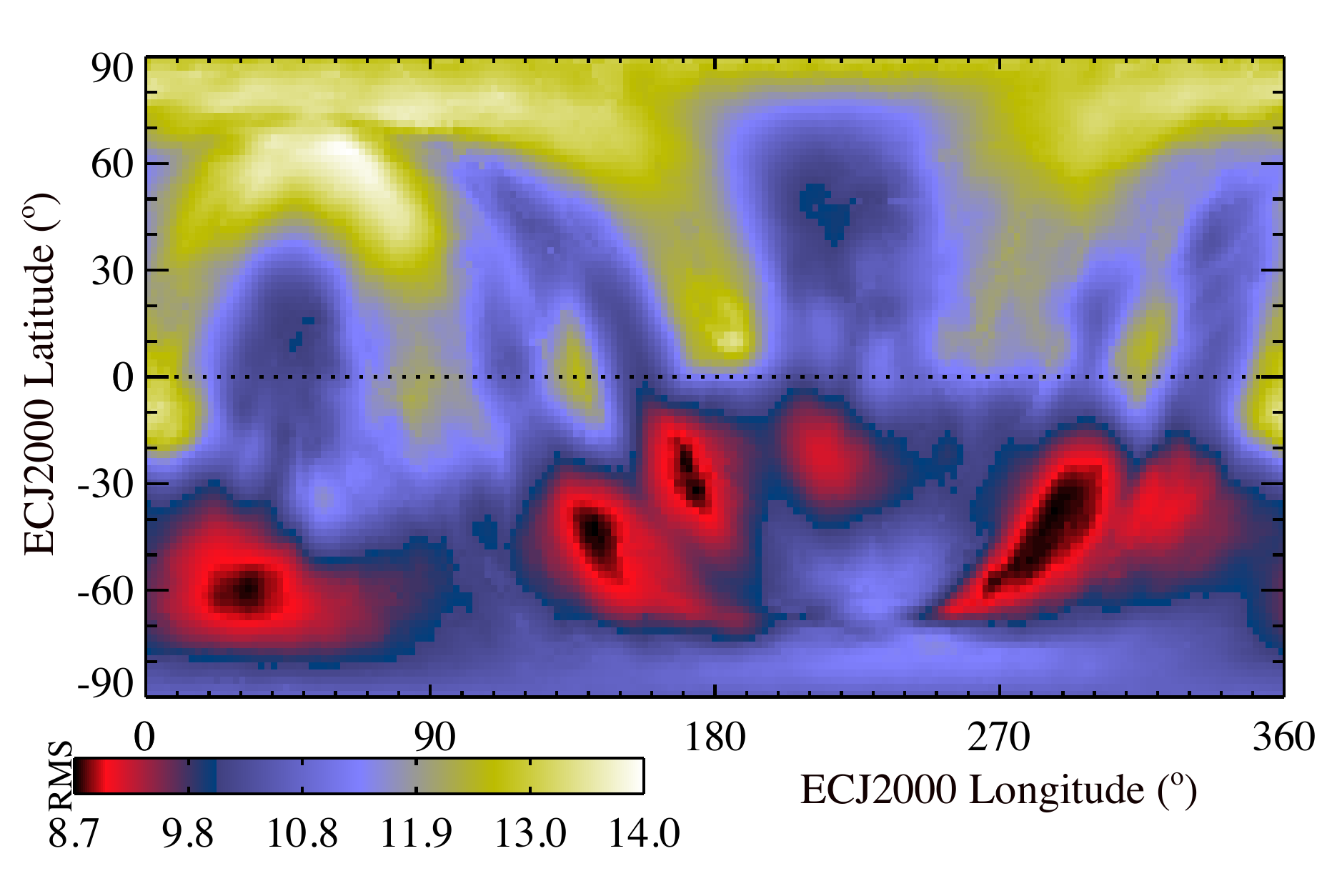}}
    \caption{The RMS deviation of the photometry relative to the
      convex model, obtained by exploring the whole space of pole
      coordinates. In the plot, the ecliptic longitude $\lambda$ is on
      the horizontal axis, and the ecliptic latitude $\beta$ on the
      vertical. Red and black values visualize the position of the
      minima.
        } 
    \label{F:map}
  \end{figure}

In running the pure photometric inversion, we decided to include
  sparse photometry to better constrain rotation period and
  spin vector coordinates, to compensate the short coverage of
    lightcurves induced by the long rotation period of Barbara. The whole procedure is described in detail in
  \citep{Hanusetal2013}. In our case we selected 182 and 124 photometric
  measurements coming from the USNO--Flagstaff station (IAU code 689) and the Catalina Sky Survey \citep[IAU code 703,][]{Larson2003}, respectively, that were added to our collection of 57 dense lightcurves.
We started by searching for optimum sidereal rotation period,
    by using the \texttt{period\_scan} software,
    available on DAMIT\footnote{\href{http://astro.troja.mff.cuni.cz/projects/asteroids3D/web.php}%
              {http://astro.troja.mff.cuni.cz/projects/asteroids3D/}}
              \citep{2010-AA-513-Durech},
    on the combined data set of
    lightcurves and sparse photometry (Fig.~\ref{F:period}).
    Starting from the best-fit period of 26.4744\,h, we then explored
    the possible locations of the spin-vector coordinates. 
    We ran a full exploration of the ecliptic J2000 celestial sphere
    ($\lambda$,$\beta$) by keeping the period fixed.
    For each position of the spin axis we computed the residuals
    of the model-derived brigthness, with respect to the photometric
    measurements. The resulting map of residuals is shown in
    Fig.~\ref{F:map}. 
    The resulting pole coordinates and rotation periods are in
    Table~\ref{T:modelparams}. 
    
  \begin{table}\scriptsize
    \centering
    \caption[]{
      Shape and spin vectors for  (234) Barbara.\label{T:modelparams} In the bottom section,
      the overall shape parameters (axis ratios of the best-fitting ellipsoid) are listed.
    }

    \begin{tabular}{lccc}
      \hline                                                                                                       
      Parameter & light-curve only & KOALA & Unit\\
      \hline  
      Period & 24.4744 $\pm$ 0.0001 & 24.4744 $\pm$ 0.0001 & h\\
      Pole ($\lambda$,$\beta$) & (156,-46) & (144,-38) & deg.\\
      \hline		  	    
      $D_V$ & & 46.3 $\pm$ 5 & km\\
      a/b   & 1.12 & 1.11 \\
      b/c   & 1.59 & 1.14 \\
      \hline		  	    
    \end{tabular}			                	           
  \end{table}

  At negative latitudes, several minima in the residual map
  appear. The two deepest ones (around $\lambda \sim$ 150\degr and
  290\degr) are both candidates for the spin axis direction.
  The fact that
  the solution is not unique clearly illustrates the challenge of the
  inversion process, despite the large number of lightcurves
  available. This is most probably due to the rather slow rotation. In
  such conditions, all the lightcurves embrace only very limited
  portion of the entire rotation (less than 20\% on average).  

  As the convex solution is not perfectly constrained, the computation
  of a shape without the convexity constraint (i.e., with concavities)
  might appear as an academic exercise. However, the concavities
  explored by the occultations are well constrained by the accuracy
  and the consistency of the timings, in particular for the event of
  November 21, 2009. For this reason, it seemed appropriate to attempt
  an inversion by KOALA, including both occultations and
  photometry. In the process, KOALA adapts the concavities to the
  occultations, but will also tend to create other less-constrained
  concavities to better reproduce the photometry. For this reason the
  result should not be taken at face value. Nevertheless we expect the
  shape to be approximately consistent with the possible VLTI
  detection of a very elongated or bi-lobated objects (see further
  below). 

  Considering the amount and the high quality of the timings
  for the stellar occultation of 2009 November 21, we chose to use the profile
  drawn by the chords as if it was obtained by disk-resolved imaging.
  This assumption was required to model the large concavity revealed
  by the interrupted chords at the South limb.   We consider that this
  approach is fully justified and does not introduce a significant
  bias, since the number of positive   chords available allowed a
  profile sampling with a resolution close to that obtained in
  disk-resolved imaging.

\begin{figure*}
    \centerline{\includegraphics[width=0.8\textwidth]{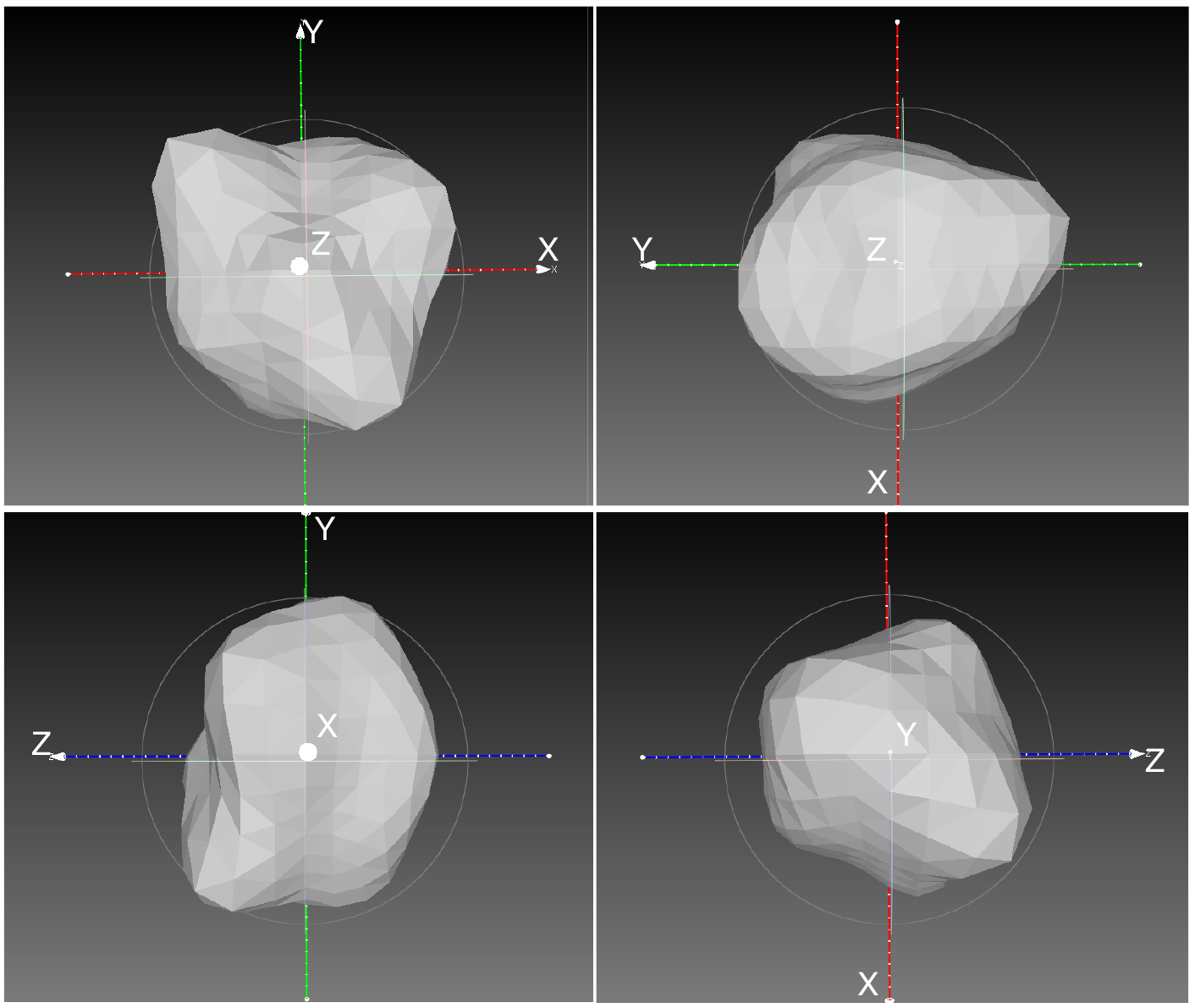}}
    \caption{Four different projections of (234) Barbara. The two
      panels at the top have the South axis pointing upward. At bottom
      left, the South polar view is presented. At bottom right, a view
      at an intermediate aspect angle, strongly enhancing the
      visibility of concave areas. The z axis is parallel to the spin axis.} 
    \label{F:Projection}
  \end{figure*}

\begin{figure}
   \includegraphics[width=0.49\textwidth]{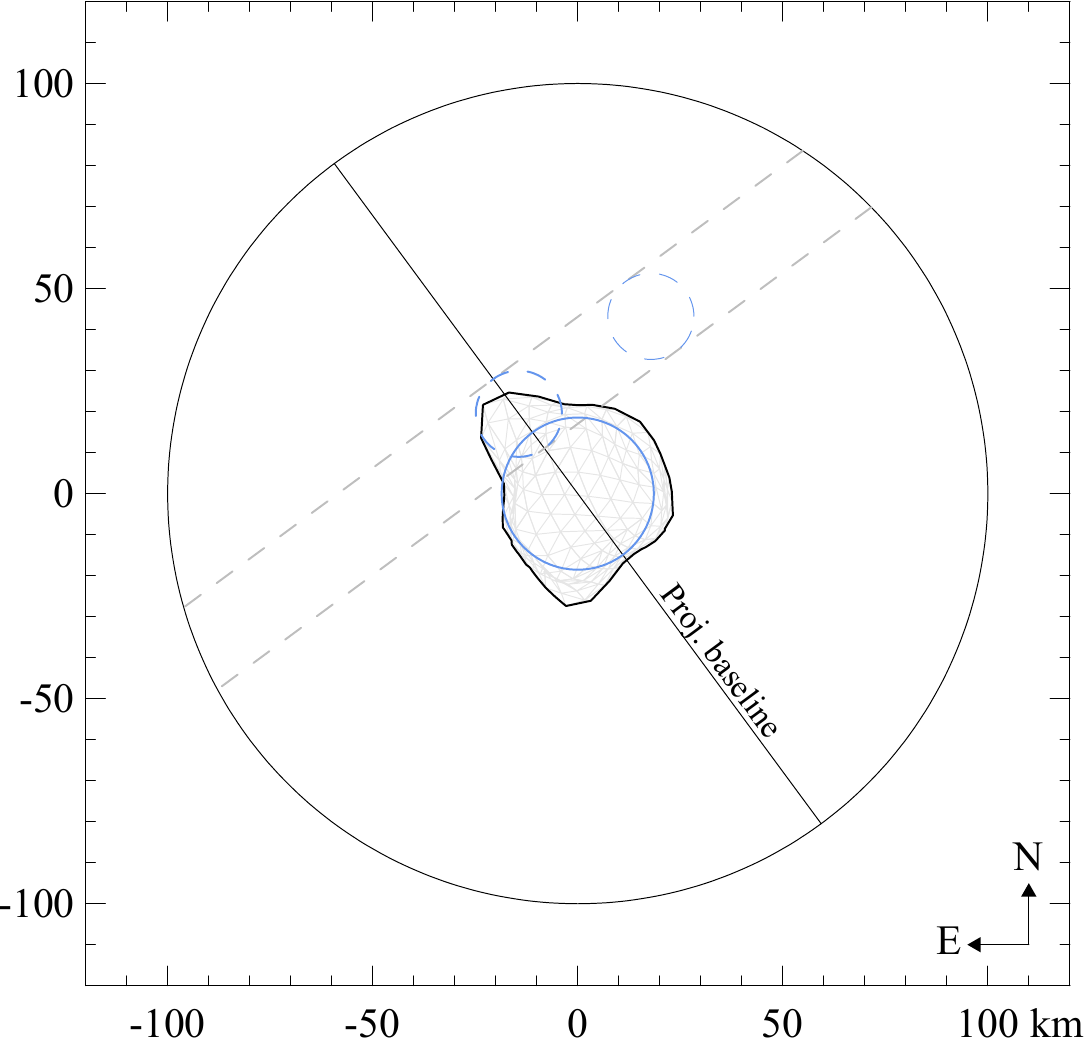}
   \caption{Comparison of the binary model of Barbara derived by
       \citet{delbo_VLTI_2009} in blue lines with the KOALA shape
       model (black contour) at the time of VLTI observations.
     The irregular shape of Barbara mislead the interpretation, by
     mimicking the signal of a binary system projected on the VLTI baseline.
     As spatial resolution is missing in the direction perpendicular to the baseline, the dashed lines
     enclose possible locations for a satellite compatible with the VLTI signal. 
   }
   \label{F:VLTI}
 \end{figure}

  We thus ran KOALA to determine the best-fit period, spin, and
    3-D shape to the lightcurves and stellar occultations, using the
    period and two spin locations determined above.
    We then checked the two solutions against the overall outline
    obtained from the MIDI-VLTI observations in 2005.
    The solution with a spin axis lying close to ECJ2000 (144\degr,
    -38\degr) provided a good match to the geometry derived from
    interferometric fitting \citep[see Fig.~\ref{F:VLTI}
      and][]{delbo_VLTI_2009}.
    The resulting shape is shown in Fig.~\ref{F:Projection}.
 
 The resulting comparison has to be interpreted by taking into account
    the orientation of the VLTI baseline, essentially aligned along the
    direction of a protruding region in the NE direction. Clearly, this
    protrusion is the most relevant irregularity found by KOALA and can
    be related to the ``secondary'' body revealed by VLTI
    observations. It also happens to be well constrained by the
    occultation data. In fact, it appears close to the main concavity
    observed in November 2009 (and pointing downward in
    Fig.~\ref{F:occ}).

  The 3-D model of Barbara derived here is made of 512
  triangular facets\footnote{The model is available on~\href{http://astro.troja.mff.cuni.cz/projects/asteroids3D/web.php}%
              {DAMIT}}
  and is scaled to absolute dimensions thanks to
  the contribution of stellar occultations. The 
  mesh volume of our model is equivalent to a sphere having a diameter of 45.9\,km.
  
  These size measurements should be compared to the thermal diameter
  derived by AKARI (47.8$\pm$0.68 km) and IRAS (43.7$\pm$1 km). WISE
  derived 53.80$\pm$1.12 km in the fully cryogenic phase
  \citep{Masiero11} and 45.29$\pm$1.33 km during the 3-band
  cryogenic/post-cryogenic operation \citep{Masiero12}. The first of
  the WISE measurements is clearly discrepant from all the other
  measurements. This discrepancy is due to the assumption made in \citet{Masiero11} that the 4.6-$\mu$m albedos are equal to the 3.4-$\mu$m albedos, which biases the relative contributions of thermal flux and reflected sunlight in the WISE 4.6-$\mu$m data. By fitting the NEATM to the fully cryogenic, purely thermal WISE bands (12 and 22 $\mu$m), we obtained a thermal diameter of 46 $\pm$ 7 km \citep[for details on our particular data selection criteria and procedure see][and references therein]{Ali-Lagoa13}. Since
the average of the four accepted values (45.7 km) is just 200\,m less than our volume-equivalent diameter (0.4\% relative difference), we consider that our model size is in excellent agreement with the thermal diameters. \\
\indent Unfortunately, the sole published mass estimate \citep{2010-SciNote-Fienga} is very poorly constrained (0.44\,$\pm$\,1.45\,$\times$\,10$^{18}$\,kg) and cannot be used to derive the density or to draw any conclusion on the internal structure. 

\section{Discussion}

  As suggested by stellar occultation and VLTI data, the shape of
  (234) Barbara is highly irregular with the presence of large
  concavities. 
  
  At all apparitions, the brightness variations seems to have approximately a similar amplitude, despite the change in aspect angle. This is probably to be ascribed to the ubiquitous irregularities - present for all illumination and observation directions.  
  
  The derived shape, when rotated at an orientation corresponding to the VLTI
  observations, suggests an interpretation of the interferometric
  signal as the presence of a big prominence oriented along the
  baseline.  
  Our results thus show that the VLTI observations can be explained without
  the presence of a large satellite. Stellar occultations also failed
  to show the presence of such a companion. Of course, this evidence
  cannot exclude that small satellites might be orbiting (234)
  Barbara (a few km in size), but they have no signatures in the available data. 
  
  Concerning the origin of the object and its collisional history, the
  absence of an identified family around (234) Barbara makes any
  interpretation loosely constrained.  

  A first possibility is that (234) Barbara is in fact the assemblage
  of two bodies of different sizes, resulting in a bi-lobated
  object. This interpretation could be consistent with the prominence
  revealed by VLTI. However, with the current limitations on the shape
  resolution and in absence of other constraints, this scenario
  remains rather speculative. 

  Another common factor of reshaping and excavation are, of course,
  non-disruptive impacts. In this case, we can reasonably assume that
  the convex hull of the shape of Barbara could represent the minimum
  volume that the original body had, before being excavated by
  impacts. If no ejecta fall back is considered, we find an
  excavated volume 6.8\% of the convex hull, representing 
  3.7$\times$10$^{12}$\ m$^3$. Since some fall-back probably occurred,
  the dislocation of the material could have been much more
  important. 

  According to \citet[][Fig. 6]{Davis02}, the probability that a 50-km
  object is a re-accumulated rubble-pile ranges from 45 to
  70\%. 
  
  The possibility of a rubble-pile structure is
  probably strengthened by the relatively slow rotation period of
  (234) Barbara. The role of non-destructive impacts on the evolution of rotation
  periods of relatively large asteroids has been the subject of several
  studies in the past. It has been suggested that collisions could
  diminish the angular momentum on average, carried away by escaping
  fragments, both in the case of craterization events \citep[``angular
    momentum drain'']{dobrovolskis_angular_1984} and in shattering
  impacts \citep[``angular momentum
    splash'']{cellino_splash_1990}. This last mechanism was invoked to
  explain the fact that asteroids smaller than $\sim$100 km have
  shorter average rotation periods than larger ones
  \citep{farinella_rotation_1992}.  
  More recently, detailed numerical simulations of impacts on
  rubble-pile asteroids \citep{Takeda2009_I, Takeda2009_II} have shown that
  spin down is in fact a common consequence of impact events. If this
  was the case for (234) Barbara, we can suggest that one or more
  impacts subtracted angular momentum from an initially much larger
  body. Internal fragmentation during the process would then be a
  natural outcome of the collisional sequence that slowed down
  rotation and excavated the large concavities.  
  A long history of collisional de-spinning and the ancient age of the 
  material seem to suggest a self-consistent scenario, supporting the idea that 
  Barbarians might be very old objects.

  The data that we have at our disposal do not show the evidence of a
  Barbara family, suggesting that the impact events could be very old,
  and the hypothetical family is now indistinguishable from the
  background. Future spectroscopic surveys could permit the detection
  of an anomalous concentrations of L/Ld-type asteroids in the corresponding region
  of the orbital elements space.
  
\section{Conclusions}

  Stellar occultations, coupled to photometry, are a powerful tool to characterize bodies that present concave features. They were seminal in
  obtaining  
  the shape of (234) Barbara, appearing to be very irregular and
  dominated by large concavities. One of these concavities is
  particularly extended and well sampled by our occultations.    

  The non-convex model that we present can still be improved and
  confirmed by more extensive observing campaigns, however it appears
  to explain the observations previously obtained at VLTI.
  The sky
  projection of the largest prominence present on the object
  is aligned to the VLTI baseline at the epoch of observation,
  mimicking the signature of a bi-lobated object. 

  Even if apparently our results seem to support the direct relation
  of concavities to polarization properties
  \citep{cellino_strange_2006}, this is valid only in the empirical
  sense, as it has no theoretical ground at present. In principle it might be
  possible that the presence of concavities is related to polarimetry
  only in an indirect sense. For example, the collisional excavation
  could have changed the composition or the texture of the surface, by
  exposing layers that remain otherwise hidden, or by
  redistributing/ejecting a layer of surface regolith. On the other hand, as the
  ``barbarians'' share common spectral properties, we cannot exclude 
  that their peculiar polarimetry is essentially due to their bulk composition. 
  The recent discovery of
  several Barbarians inside the Watsonia family seems to corroborate
  this hypothesis, as it implies the transmission by direct heritage
  of the Barbarian properties from the parent body to family members
  \citep{Cellinoetal2014}.  
  
  We underline the role played in this work by a large collaboration of
  amateur and professional astronomers, both in obtaining an adequate
  photometric coverage and in securing the positive results of the two
  occultations that have allowed us to constrain the concavities. The
  long rotation period, and its commensurability to the duration of Earth's day
  is an obstacle to an efficient coverage of the object
  rotation.  
  
  A network of observers at different longitudes constitute a clear
  advantage that we will try to exploit in future photometric
  campaigns devoted to barbarian asteroids. Obtaining more shapes of
  the Barbarians, polarimetric measurements and near-infrared spectra
  to confirm the presence of spinel-rich inclusions is required to better understand
  such objects that could represent a rare sample of CAI-rich Solar
  System bodies, dating back to the first phases of Solar System
  formation.

\section*{Acknowledgments}

  This research used the Miriade VO tool developed at IMCCE
  \citep{2008-ACM-Berthier} and MP3C at OCA (Delbo' and Tanga,
  http://mp3c.oca.eu/). We acknowledge the financial support of the
  occultation activity carried on by OCA members, from the BQR program of the
  Observatoire de la C\^ote d'Azur, the Action Specifique Gaia,
  and the Programme Nationale de Plan\'etologie.
  The work of JH was carried under the 
  contract 11-BS56-008 (SHOCKS) of the French Agence National 
  de la Recherche (ANR). We thank the developers and maintainers of Meshlab and VO
    Topcat software. TRAPPIST is a project funded by the Belgian Fund for Scientific Research (Fonds de la Recherche Scientifique, F.R.S FNRS)
under grant FRFC 2.5.594.09.F, with the participation of the Swiss National Science Fundation (SNF). E. Jehin and M. Gillon are FNRS Research Associates, J. Manfroid is Research Director FNRS. A. Matter acknowledges financial support from the Centre National d'Etudes Spatiales (CNES). We thank all the
  members of ``Agrupaci\'on Astron\'omica de Fuerteventura'' for the
  coordination of the observations and the ``Cabildo de Fuerteventura''
  for the logistic support in Tef\'ia observatory. Also, S. Degenhart
  is acknowledged for the organization of the campaign in the
  U.S.A. that yielded the second occultation profile presented in this
  paper. 


\bibliographystyle{mn2e}
\bibliography{phys_properties}

\label{lastpage}

\end{document}